\documentclass[pre]{revtex4}
\usepackage[margin=3cm,a4paper]{geometry}
\usepackage{graphicx}
\usepackage{latexsym}
\usepackage{amsmath}
\usepackage{amssymb}
\usepackage{amsfonts}
\usepackage{amsthm}
\usepackage{epstopdf}
\usepackage{epsf}
\begin{document}
\title{Dynamical study of nonlinear ion acoustic waves in presence of charged space debris at Low Earth Orbital (LEO) plasma region}
\author{Abhik Mukherjee
\footnote{Electronic mail: abhikmukherjeesinp15@gmail.com, mukerdzhi.a@misis.ru}}
\affiliation{Department of Theoretical Physics and Quantum Technologies, National University of Science and Technology, ``MISiS'', Moscow, Russia}
\author{S. P. Acharya
\footnote{Electronic mail: siba.acharya@saha.ac.in and siba.acharya39@gmail.com}}
\affiliation{Saha Institute of Nuclear Physics, Kolkata, India}
\author{M. S. Janaki
\footnote{Electronic mail: ms.janaki@saha.ac.in}}
\affiliation{Saha Institute of Nuclear Physics, Kolkata, India}

\begin{abstract}
We consider the nonlinear ion acoustic wave  induced  by the orbiting charged space debris in the plasma environment generated at Low Earth Orbital (LEO)
region. The generated nonlinear ion acoustic wave is shown to be governed by the forced Korteweg-de Vries (fKdV) equation with the forcing function dependent on the charged space debris function.
For a specific relationship between the forcing debris function and the nonlinear ion acoustic wave, the forced KdV equation turns to be a completely integrable system where the debris function obeys a definite nonholonomic constraint. A special exact accelerated
soliton solution (velocity of the soliton changes over time whereas it's amplitude remains constant) has been derived for the ion acoustic wave for the first time. On the other hand, 
the amplitude of the solitonic  debris function varies with time and it's shape changes during propagation. Approximate ion acoustic solitary wave solutions with time varying amplitude and velocity, have been derived for different  weak localized charged debris functions. Possible applications of the obtained results in space plasma physics are stated along with future direction of research.
\end{abstract}

\maketitle

\section{Introduction} \label{intro}
Research in space plasma physics and technologies increased drastically after the launch of Sputnik-1 \cite{Sputnik}, the first artificial Earth satellite, on 4 October 1957 and still remains one of the important research fields
in plasma science and technology.
In recent times, the near-Earth space has become a laboratory for studying plasma physics.
Scientists have sent rockets and satellites to the ionosphere \cite{Kelley} to gather scientific data because it becomes impossible to create on  earth a laboratory as vast and variable as the ionosphere. 
In this regard, the research on the dynamics of space debris \cite{Klinkrad} has gained its importance.
The inactive space objects like dead satellites, destroyed spacecrafts, meteoroids and other
inactive materials resulting from many natural phenomena are collectively called ``space debris'' \cite{Klinkrad}. Their
 sizes vary from  microns to centimeters depending on the conditions \cite{NASATR}. These are found substantially in the Low Earth Orbital (LEO) and Geosynchronous Earth Orbital 
(GEO) region with high velocities like 10km/s \cite{ASA}. As these objects are immersed in a plasma medium and exposed to many interplanetary radiations including solar radiations, 
they get charged due to photo emission, electron and ion collection,
secondary electron emission \cite{Horanyi} etc. This results in a significant effect on other  space objects \cite{Paul, Juhasz}. Due to their high 
velocities particularly in the GEO region, even micron-sized objects can affect the running spacecrafts a lot \cite{NASATR}. Thus,
in order to get rid of these deteriorating effects, the dynamical study of charged space debris is becoming important.
The active debris removal (ADR) becomes a challenging problem in the twenty-first century due to their enormous number in a complicated plasma environment particularly  in
 the GEO region \cite{Anderson}. Recently many attempts have been made for the ADR in the LEO region like electrodynamics tethers for the LEO applications \cite{Pearson}, electric sails \cite{Janhunen}, 
 space debris pushers \cite{Bombardelli}, robotic docking \cite{Smith} etc.  Various explorations on tracking the complex network of the space debris orbits have also been done by many space agencies
 in order to look for admissible orbits and prevent collisions between the spacecrafts and debris objects. 
 There also exist many indirect tracking methods for the debris problem. 
Recently, a new detection method of charged debris objects in LEO region has been discussed by Sen et al \cite{Sen} where the generated nonlinear ion acoustic solitons are modelled by forced KdV (fKdV) equation. They 
have discussed few exact line solitons of  the equation where the debris field also have the localized form. They have carried out a numerical computation taking the forcing term as Gaussian function to show   ``precursor
solitons''. Such solitons which are emitted periodically  can be detected and give an indirect evidence of the existence of charged debris \cite{Sen}.
Many studies on the fKdV equation have been carried out in plasmas considering periodic forms of the source function
\cite{TFAli,TFChatterjee,TFZhen}. In superthermal plasmas \cite{TFChatterjee},
 plasmas with $\kappa$
distributed particles \cite{TFAli} the solitary wave structure of fKdV equation has been studied. In Thomas-Fermi plasmas,  periodic, quasiperiodic and chaotic structures in the presence of sinusoidal
source debris term are also explored \cite{TFMandi}. But the debris function can also have many other kinds of localized forms  which are not studied yet.
All these studies till now discuss either numerical or perturbative solitary wave solutions. Exact solutions are discussed in \cite{Sen} which are line solitons with constant amplitude and velocity.
But in real physical situations, the velocities of both ion acoustic wave and the source debris term may vary with time showing accelerating features. Again, since source debris function induces such variations
in nonlinear ion acoustic waves, hence they may be self consistently related to each other.   Considering these possibilities, 
a thorough dynamical study of the nonlinear ion acoustic waves in presence of charged debris function is necessary discussing exact  accelerated solitons, different kinds of exact as well as approximate solitary wave solutions etc. 
In this work, we will discuss  about a special kind of exact  soliton solution, the velocity of which changes over time instead of being constant like line soliton. It is generated  self consistently by the shape changing debris soliton
which is not studied till now as far as our knowledge goes. 
For a specific dependence between the debris function and the ion acoustic wave, the forced nonlinear evolution equation becomes a completely integrable system. The detailed mathematical theory of such kind of forced yet integrable equations are discussed in \cite{Kundu,Kundu1}. 
The nonlinear  ion acoustic soliton accelerates depending on the debris field whereas it's amplitude remains constant. Since, it is an exact solution of a completely integrable system obeying elastic scattering property \cite{Kundu}, it can be called as ``soliton'' in spite of having it's varying velocity. On the other hand, both 
amplitude and velocity of the solitonic debris function changes over time. For  general weak localized debris functions, the approximate solitary wave solutions with varying amplitude and velocity, can be obtained for the ion acoustic wave using an approximation technique.

The paper is organized as follows. The derivation of the nonlinear evolution equation for the ion acoustic wave in presence of such charged debris field is shown in section-II. The condition of integrability of the forced evolution equation is discussed in section-III along with the special exact  soliton solutions having accelerating features (the velocity varies with time whereas the amplitude remains constant).  In section-IV, we discuss some  approximate solitary wave solutions for various functional forms of the localized debris field.  The discussions of our obtained results along with possible applications are given in section - V. Conclusive remarks  are stated in section-VI followed by 
acknowledgments and bibliography.

 \section{Derivation of the  evolution equation for the nonlinear ion acoustic waves in presence of charged space debris} \label{evolution}
We consider, the propagation of long wavelength - finite amplitude nonlinear ion acoustic waves in the Low Earth Orbital (LEO) region 
due to the motion of orbital charged debris.  The region which we consider is populated by a low temperature low density plasma. We consider the propagation of nonlinear
ion acoustic waves, where the ion species is treated as the cold species i.e. the ion pressure is neglected and the electrons are assumed 
to obey a Boltzmann distribution. The basic normalized system of equations in this system in (1+1) dimensions are given by : 
\begin{equation} 
 \mathrm{\frac{\partial n}{\partial t}+\frac{\partial (nu)}{\partial x}=0},\label{cont}
 \end{equation}
 \begin{equation}
\mathrm{\frac{\partial u}{\partial t}+u\frac{\partial u}{\partial x}+\frac{\partial \phi}{\partial x}=0}, \label{momt} 
\end{equation}
\begin{equation}
\mathrm{\frac{{\partial}^2 \phi}{\partial x^2}-e^{\phi}+n= S(x,t)} \label{Posn},
\end{equation}
 where the following normalizations have been used:
  $$\mathrm{x\longrightarrow x/{\lambda}_d; \,t\longrightarrow \frac{C_s}{{\lambda}_d}t; \, n\longrightarrow \frac{n}{n_{0i}};\, u\longrightarrow \frac{u}{C_s};\, \phi \longrightarrow \frac{e \phi }{k_BT_e}}$$
  where ${\lambda}_d$ is  electron Debye length, $\mathrm{C_s}$ is  ion acoustic speed, $\mathrm{k_B}$ is  Boltzmann constant, $T_e$ is electron temperature and $n_{0i}$ is  equilibrium ion density. Equations (\ref{cont}), (\ref{momt}) and (\ref{Posn}) represent  ion continuity equation, ion momentum conservation equation and Poisson's equation respectively, where
  n, u,  $\phi$ denote the density, velocity of the ion species and the electrostatic potential respectively. The
  term : $\mathrm{S(x, t)}$ in the RHS of equation (\ref{Posn}),  represents a charge density source arising due to the
  passage of the charged debris. 
  We introduce two stretched variables as:
 \begin{equation}
 \mathrm{\xi={\epsilon}^{1/2}(x- v_p t)}; \,    \tau ={\epsilon}^{3/2}t,
 \end{equation}
 where, $v_p$ is a constant and $\epsilon$ is a small perturbation parameter which we will use later.
 In the paper \cite{Sen}, Sen et. al have considered the analytic ion acoustic wave  solution and the source term $\mathrm{S}$ to have the form of  line solitary waves, where the amplitude and velocity are constants. 
  But in this work we do not follow such restrictions, because in reality such parameters of both the debris source term and the ion acoustic wave may vary with time. Also such variations
  of the source debris wave and nonlinear ion acoustic wave may be self consistently related to each other since the charged debris can induce such nonlinear excitations.
  We derive briefly the evolution equation for the nonlinear ion acoustic wave in the presence of source term $\mathrm{S}$ using reductive perturbation technique (RPT) \cite{RPT_arxiv}. 
 Each dependent variable is expanded in series of a small parameter $\epsilon$ as:
  \begin{equation} 
 \mathrm{n= 1+ \epsilon n_1+{\epsilon}^2n_2+ O(\epsilon^3)},
\end{equation}
\begin{equation} 
 \mathrm{u=\epsilon u_1+{\epsilon}^2u_2+ O(\epsilon^3)},
\end{equation}
\begin{equation} 
 \mathrm{\phi=\epsilon {\phi}_1+{\epsilon}^2{\phi}_2+ O(\epsilon^3)}.
 \end{equation}
We consider a weak space-time ($\xi - \tau$) dependent localized debris function which vanishes at space infinities. Hence, after scaling we get the following form :
  \begin{equation}
   S = \epsilon^2 f (\xi,\tau) \label{S},
  \end{equation}
  where $f (\xi,\tau)$ can have any spatially localized form that is consistent with the debris dynamics.
Following the steps of the perturbative calculation given in \cite{Sen} we can arrive at the equation :
\begin{equation}
n_{1\tau}+n_1 n_{1\xi} + \frac{1}{2} n_{1 \xi\xi\xi} =  \frac{1}{2} f_{\xi}, \label{fKdV}
\end{equation}
which is nothing but the forced Korteweg-de Vries (fKdV) equation, where subscripts denote partial derivatives. Changing to a new set of variables : $n_1=3U, \tau=2T, \xi=X, F = \frac{1}{3} f $, we get
a more convenient form as
\begin{equation}
U_{T} + 6 \ U \ U_{X} +  U_{XXX} =  F_{X}.  \label{sfKdV}
\end{equation}
Thus we get a forced KdV equation (\ref{sfKdV}) as the evolution equation for the nonlinear ion acoustic wave in presence of charged debris term. We know that Eq.  (\ref{sfKdV}) is not exactly solvable in general but for a specific relation between $U$ and $F$,  it  turns into a completely integrable system which is discussed in detail in the next section. We will also show that Eq.  (\ref{sfKdV}) under that integrable condition, admits a special exact accelerated soliton solution, the velocity of which varies with time whereas the amplitude remains constant.

\section{Condition of integrability of the evolution equation  (\ref{sfKdV}) : some novel properties}
As far as we know, there is no generalized exact solution of  the forced KdV (fKdV) equation (\ref{sfKdV}) in the scientific literature. The simple reason is that the forcing function destroys the complete integrability of the KdV equation, hence it's exact solvability is also lost.
Using Gaussian debris function with velocity $v_d$, the fKdV equation is solved numerically in \cite{Sen} and the generation of pre-cursor solitons is explained as well.
Such solitons which are emitted periodically  can be detected and give an indirect evidence of the existence of charged debris \cite{Sen}.
Sen et al. \cite{Sen} have discussed  few exact  line solitary waves of the forced KdV 
equation taking the localized forcing functions. Many studies  have been done  in this field considering periodic form of debris function \cite{TFAli,TFChatterjee,TFZhen}. 
All these studies till now discuss either approximate or numerical solutions. 
But in actual perturbative condition, the velocities of both ion acoustic soliton and debris soliton may vary with time. Again the ion acoustic wave and the debris field may be self consistently related to each other. In this section,  we will derive a special  exact  soliton solution for ion acoustic wave as well as the debris field, the velocity of which varies with time. This type of exact localized solutions with unusual features are not studied in this field  as far as our knowledge goes.
Also, the acceleration associated with such novel exact solutions are arbitrary. Hence choosing the arbitrary function accordingly, we can model the real wave phenomena.

It should be noted that nonholonomic deformation theory has been applied to KdV equation preserving its integrability in \cite{6KdV}. In that case, the forcing function present in fKdV equation satisfies a nonholonomic constraint so that the equation becomes completely integrable. The resulting equation can be transformed into integrable sixth-order KdV equation using that nonholonomic constraint. The properties of complete integrability of this KdV equation with nonholonomic constraint i.e, existence of infinite number of conserved quantities, Lax pair, N soliton solutions, solvability via inverse scattering transform method, elastic nature of soliton scattering, two fold integrable hierarchy  etc are discussed in detail in \cite{Nonhol,Kundu}. Infinite number of generalized symmetries and a recursion operator has been constructed for this integrable system  in \cite{Kundu1}. The nonholonomic deformation of KdV equation has been shown to be of the form of a self  consistent source equation (SCSE) having particular exact solutions in \cite{Yao}. On the other hand the concept of SCSE
has been shown in \cite{Melnikov} where the author discusses a coupled system consisting the main integrable equation with an extra term which is made from the eigenfunctions and  eigenvalue equations of the Lax operator.

Thus we can state as the gist of the above established theories, that our fKdV equation
can be transformed into a completely integrable system if the forcing debris function $F$ satisfies a nonholonomic contraint. Following the same mathematical form that is used in \cite{Kundu,Kundu1}, we make a simple transformation as
(\ref{sfKdV}) 
by $X \longrightarrow - X, F \longrightarrow - F$. Hence, equation (\ref{sfKdV}) turns into a completely integrable equation as  
\begin{equation}
U_{T} - 6 \ U \ U_{X} -  U_{XXX} =  F_{X},  \label{ffKdV}
\end{equation}
where the forcing function, $F$ which is present in (\ref{ffKdV}) follows the following constraint : 
\begin{equation}
 F_{XXX} + 4 U F_X + 2 U_X F = - 2 a(T) U_X, \label{NHC}
\end{equation}
where $a(T)$ is an arbitrary function of $T$. The significance of $a(T)$ will be analyzed later.
The system of equations (\ref{ffKdV}) and (\ref{NHC})  constitute an integrable system having a special exact soliton solution with varying velocity. They are called ``solitons'' \cite{Kundu,Kundu1} in spite of having varying velocity because they are the solutions of a completely integrable system (Eq. (\ref{ffKdV}) and (\ref{NHC})) and follow elastic soliton scattering property \cite{Kundu}. We can obtain integrable sixth order KdV equation discussed in \cite{6KdV} from (\ref{ffKdV}) and (\ref{NHC}) by using a simple transformation of variables.
 In this section, we will discuss  certain interesting properties of our integrable system  (\ref{ffKdV}) and (\ref{NHC}) like special accelerated soliton solutions,  Lax pair, infinite number of conserved quantities, higher soliton solutions etc to describe it's mathematical features.

 \subsection{Exact accelerated solitons} \label{soliton}
We will discuss about  a special exact one soliton solution with varying velocity \cite{Kundu,Kundu1} for both the ion acoustic wave $U$ and the debris field $F$ from  Eqns (\ref{ffKdV}) and (\ref{NHC}).
 The special one soliton solution of the ion acoustic wave $U(X,T)$ is given by
\begin{eqnarray}
 U(X,T) = 2 k^2 sech^2{\eta}, \ \  
 \eta = k \{ X + V(T) T \} + \eta_1, \ \ V(T) = 4 k^2 - \frac{\int a(T) dT}{2 k^2 T}, \label{Accsol}
\end{eqnarray}
where $k, \eta_1$ are constants. We can see from the solution (\ref{Accsol})
that the velocity $V(T)$ of the soliton varies with time showing acceleration due to the presence of the arbitrary function $a(T)$ whereas the amplitude remains constant.
On the other hand one soliton solution for the debris field $F$ is given by
\begin{eqnarray}
 F(X,T) = - a(T) sech^2{\eta}, \ \  
 \eta = k \{ X + V(T) T \} + \eta_1, \ \ V(T) = 4 k^2 - \frac{\int a(T) dT}{2 k^2 T}, \label{debsol}
\end{eqnarray}
showing variation in both amplitude and velocity. Again as $X \longrightarrow \pm \infty$, the function $F$ vanishes satisfying (\ref{Posn}).
The dynamical features of these solutions (\ref{Accsol}) and (\ref{debsol}) for different $a(T)$ are shown in FIG. 1, 2.
Thus, the special exact accelerated one soliton solution for the ion acoustic wave $U$ in presence of charged debris term $F$ is derived for the first time as far as our knowledge goes. On the other hand, the acceleration of the soliton and the debris field is arbitrary since, $a(T)$  is arbitrary.
Similarly, the higher order exact accelerating solitons can also be derived which are shown later. The velocity of the ion acoustic soliton
changes over time showing accelerating/decelerating features whereas its amplitude remains constant. Whereas both the amplitude and velocity of the source debris term,
which has also solitonic nature, changes over time showing shape changing effects.  
Different shapes of $U, F$ can be obtained by choosing different functional forms of the arbitrary function $a(T)$. 
One important point should be noted that such accelerated  soliton solutions can only be obtained when the forcing debris term $F$ of (\ref{ffKdV}) satisfies the constraint (\ref{NHC}) forming an integrable system. 
For any other conditions, equation (\ref{ffKdV}) looses its integrable  feature allowing approximate solitary wave solutions as discussed in section-IV. We can see from (\ref{Accsol}), (\ref{debsol}) that the only difference in the functional forms of $U$ and $F$ is  in their expression of amplitudes.  So the ion acoustic wave $U$ is generated self consistently by the debris $F$ having similar dynamic velocity profile. But the amplitude of $U$ remains constant whereas that of $F$ varies with time changing it's shape. Since, the analytic forms of the time varying velocities of both $U$ and $F$ are identical, they move together. So, they can be regarded as ``pinned solitons'' as discussed in \cite{Sen}.

\begin{figure}

\includegraphics[width=8cm]{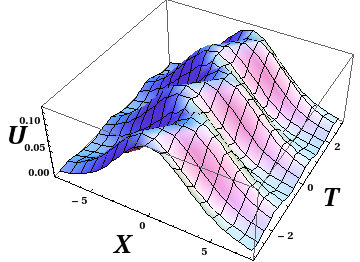}
 
\vspace{1cm}

 (a) Ion acoustic one soliton solution $U$ for $a(T) = 0.5 \cos(3T)$.

\vspace{1cm}

\includegraphics[width=8cm]{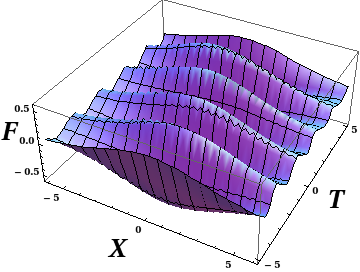}

\vspace{1cm}

(b) One debris soliton solution $F$
for $a(T) = 0.5 \cos(3T)$.

\vspace{1cm}

\noindent FIG.1: {\bf 
 The trigonometric  form of $a(T)$ causes the phase of the ion acoustic solitary wave to change periodically which causes acceleration
in $X,T$ plane, whereas it's amplitude remains constant. One the other hand both amplitude and velocity of the source debris soliton $F$ varies with time. This change in amplitude
of the charged space debris may be may be detected and can provide an indirect evidence it's existence. The constants are taken as $k=0.25, \eta_1 = 0$.}
\end{figure}

\begin{figure}

\includegraphics[width=8cm]{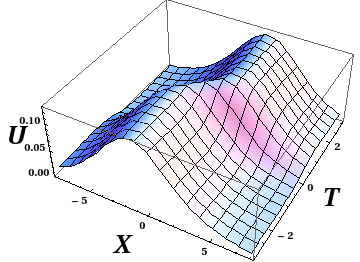}
 
\vspace{1cm}

 (a) Ion acoustic one soliton solution $U$ for $a(T) = 0.5 sech(3T)$.

\vspace{1cm}

\includegraphics[width=8cm]{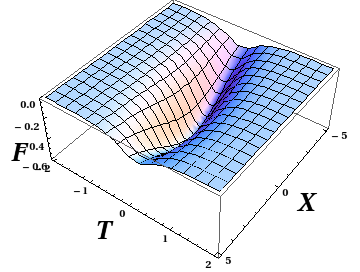}

\vspace{1cm}

(b) One debris soliton solution $F$
for $a(T) = 0.5 sech(3T)$.

\vspace{1cm}

\noindent FIG.2: {\bf 
 The hyperbolic  form of $a(T)$ causes the phase of the ion acoustic solitary wave to change showing bending 
in $X,T$ plane, whereas it's amplitude remains constant. One the other hand both amplitude and velocity of the source debris soliton $F$ varies with time. This change in amplitude
of the charged space debris may be may be detected and can provide an indirect evidence it's existence. The constants are taken as $k=0.25, \eta_1 = 0$.}
\end{figure}

 \subsection{Lax pair}
 Existence of Lax pair operators is one of the necessary conditions of an integrable system. 
 Our completely integrable system (\ref{ffKdV}) and (\ref{NHC}) have the following AKNS-type matrix Lax pair $M(\lambda),N(\lambda)$  satisfying the flatness condition
 \begin{equation}
  M_T - N_X +[M,N] = 0, \label{fl}
 \end{equation}
where $\lambda$ is the spectral parameter \cite{Kundu}.
Lax operators $M$ and $N ( = N_{KdV} + N_D$) are given by,
\begin{eqnarray}
M = 
\begin{bmatrix}
i\lambda & iU  \\
i & -i\lambda
\end{bmatrix}, \ \ 
N_{KdV} = 
\begin{bmatrix}
(-4i\lambda^3+2iU\lambda+U_X) & (-4iU\lambda^2-2U_{X}\lambda+iU_{XX}+2iU^2)  \\
(-4i\lambda^2+2iU) & (4i\lambda^3-2iU\lambda-U_X)
\end{bmatrix},
\end{eqnarray}
\begin{eqnarray}
N_D = \frac{1}{2\lambda}
\begin{bmatrix}
\{\frac{F_X}{2\lambda}+i(F + a)\} & (-F_X-\frac{e}{\lambda})  \\
\frac{i(F+ a)}{\lambda} & \{-\frac{F_X}{2\lambda}-i(F+ a)\}
\end{bmatrix} \label{LaxKdV}, \ \ e_X = i U F_X.
\end{eqnarray}
It can be checked that if these Lax operators $M$ and $N$  are inserted in the flatness condition (\ref{fl}) then it will give Eq. (\ref{ffKdV})
along with the nonholonomic constraint (\ref{NHC}).
  Details of such integrable nonholonomic deformations are discussed in \cite{Kundu1,Kundu}.

\subsection{Conserved quantities}
The local conservation law of a (1+1) dimensional partial differential equation takes the form
\begin{equation}
 \frac{\partial \rho}{\partial T} + \frac{\partial J}{\partial X} = 0, \label{cl}
\end{equation}
which has to be satisfied by all solutions. The quantities $\rho(X,T)$ and $J(X,T)$ are called local conserved density and current density respectively. We can show that our system (\ref{ffKdV}) and (\ref{NHC})
has infinitely many conserved quantities $C_n$ which is a signature of completely integrable system. 
Using (\ref{ffKdV}) and (\ref{NHC}), we can derive the first conservation law which is nothing but conservation of mass where
\begin{equation}
 \rho_1 = U, \  J_1 = - U_{XX} - 3U^2 - F.
\end{equation}
Similarly, the second conservation law (conservation of momentum) can be derived as :
\begin{equation}
 \rho_2 = U^2, \  J_2 = -4U^3 - 2UU_{XX} + U_X^2 + 2UF + F_{XX} + 2 a(T) U.
\end{equation}
Similarly we can derive infinitely many conserved densities  of this system.
 We can see from the above analysis that the conserved 
densities remain the same as in case of unforced KdV equation whereas the current densities explicitly contain the forcing debris function $F$. One important thing must be stated 
from Liouville integrability \cite{IST} that
the  conserved quantities $C_n, n=1,2,...,$ must commute i.e, $\{C_m,C_n\}$=0, as a necessary condition. From (\ref{cl}), we can see that
if the wave profile $U$ and its derivatives vanish at space infinity, we can evaluate conserved quantities as 
\begin{equation}
 C_1 =\int UdX, \ C_2 = \int U^2 dX, \ C_3 = \int (4U^3 - 2U_X^2)dX,
\end{equation}
and so on where the limit of the integration is taken from $-\infty$ to $+\infty$. 
A more mathematically detailed way to find conserved quantities of integrable equations can be found from Inverse scattering transform analysis. One can systematically 
evaluate them using the space Lax operator in Ricatti equation which is discussed in detail in \cite{IST}.

\subsection{Higher soliton solutions}
Being an integrable system, equations (\ref{ffKdV}) and (\ref{NHC}) can be solved to produce exact higher soliton solutions via many analytic mathods like Inverse scattering transform, Hirota bilinearization technique etc.
We have discussed before about the special one soliton solution  which shows acceleration instead of having constant velocity, due to presence of the arbitrary function $a(T)$ in it's argument.
In the same way we can find  exact two soliton solution of the ion acoustic wave as 
\begin{eqnarray}
 U =\frac{2}{D^2}(D_{XX}D-D_X^2), \ \  D = 1 + e^{-\eta_1} + e^{-\eta_2} + p_{12} e^{-(\eta_1 + \eta_2)} \nonumber \\ 
 \eta_1 = 2 k_1(X + V_1(T) T) + \eta_1^{0},  \ \
 \eta_2 = 2 k_2(X + V_2(T) T) + \eta_2^{0} , \ \ p_{12} = (\frac{k_1-k_2}{k_1+k_2})^2, \nonumber \\
 V_{1,2}(T) = V_{01,02} + V_{d1,d2}(T) , \ \ V_{01,02}=4k_{1,2}^2 , \ \ V_{d1,d2}(T)= \frac{-2\int a(T)dT}{(V_{01,02}T)},\label{2soliton}
\end{eqnarray}
 where $\eta_{1,2}^0, k_{1,2}$
are constant parameters. Hence, the velocity of the two soliton solution  (\ref{2soliton}) changes (showing acceleration) due to the presence of the time varying function $a(T)$ in $V_{1,2}$ that can be seen from FIG. 3.
In the same way, we can find recursively the $N$ soliton solution of our system which we do not show here due to notational complexity.
On the other hand two soliton solution for the debris function $F$ is given as :
\begin{eqnarray}
 F =\frac{2a(T)}{D^2}(D\tilde{D_{X}}-\tilde{D} D_X), \ \  \tilde{D} = \frac{1}{k_1} e^{-\eta_1} +\frac{1}{k_2} e^{-\eta_2} + p_{12}(\frac{1}{k_1}+\frac{1}{k_2}) e^{-(\eta_1 + \eta_2)}.
 \label{2solitonD}
\end{eqnarray}
It can also be proved that this integrable system  has infinitely many generalized symmetries and a recursion operator which are the signatures of integrability \cite{Kundu1}.

\begin{figure}

\includegraphics[width=8cm]{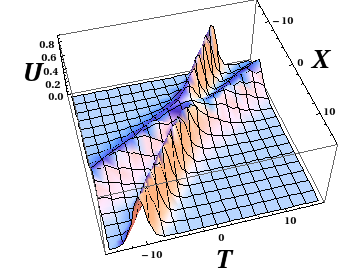}
 
\vspace{1cm}

 (a) Ion acoustic two soliton solution $U$ without the presence of debris field i.e,  for $a(T) = 0$.

\vspace{1cm}

\includegraphics[width=8cm]{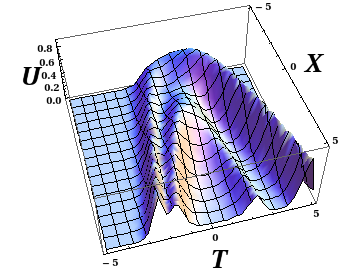}

\vspace{1cm}

(b) Ion acoustic two soliton solution $U$ with the presence of debris field i.e,  for $a(T) = T $.

\vspace{1cm}

\noindent FIG.3 : {\bf 
 The presence of the debris term $a(T)$ causes the phase of the ion acoustic two soliton solution (\ref{2soliton}) to change showing bending 
in $X-T$ plane, whereas it's amplitude remains constant. The constants are taken as $k_1 = 0.6,
\eta_1^{0} = 0.5,
k_2 = 0.4, 
\eta_2^{0} = 0.7$.}
\end{figure}

 \section{ analytic solitary wave solutions for the  ion acoustic waves for various localized charged debris functions }
In  the previous section, we have discussed about the properties of the integrable equations (\ref{ffKdV}), (\ref{NHC})   possessing  
 exact accelerated soliton solutions. It is quite surprising to see that though there exists a forcing function  yet for a specific condition the fKdV equation (\ref{ffKdV}) turns into a completely integrable system. Though it is very interesting mathematically, yet it's physical significance may not be so strong because the debris term may not always satisfy such constraint (\ref{NHC}). Hence, we need to study the behavior of ion acoustic solitary waves using approximation technique for other choices of localized debris functions.
 In this section, we will derive the approximate ion acoustic solitary wave solutions for various localized small debris fields following the approximation method described in \cite{Ghosh, OttSudan}.  The amplitude of the ion acoustic solitary wave varies very slowly with time depending on the small parameter present in the debris function.
We multiply our original  forced KdV equation (\ref{sfKdV}) by $U$ and integrate over $X$ from $-\infty$ to $+\infty$ to get
\begin{equation}
\frac{\partial}{\partial T} \int_{-\infty}^{\infty}  U^2 dX
= \int_{-\infty}^{\infty} 2 U F_X  dX, \label{condition}
\end{equation}
where we have assumed that both $U$ and it's spatial derivatives vanish for $X \longrightarrow \pm \infty$.
We assume a small debris function $F$  such  that $ F(X,T) = \delta f_1(X,T)$, where $\delta$ is a small positive parameter.
 Then, following \cite{Ghosh}, we  assume the form of the solution $U$ as
\begin{equation}
 U(X, T) = {A(T)} \ sech^2 \ [\sqrt{\frac{A(T)}{2}}\{ X - 2 \ A(T) \ T \} ],\label{psol}
\end{equation}
where the amplitude $A(T)$ varies slowly with time depending on the small parameter $\delta$. Now we will consider various spatially localized charged debris functions and calculate the analytical variation of the amplitude of the ion acoustic solitary waves with time. 
In earlier scientific literature in this field, only periodic debris functions are considered 
\cite{TFAli,TFChatterjee,TFZhen}. Hence,  other possible localized forms of debris functions should be considered in order to study the dynamical behavior of ion acoustic solitary waves in such environment.

In section III, we have considered a self consistent debris function which affects the ion acoustic wave self-consistently and vice versa. When the debris field $F$ obeys a constraint (\ref{NHC}), the system becomes completely integrable having exact accelerating solitons. For other choices of the debris function, the integrability of the system gets lost. Then using a definite perturbation scheme \cite{Ghosh} for the wave dependent source, we can obtain the analytic expressions of the ion acoustic solitary waves. In such cases, i.e., when $f_1 = f_1  (U, U_X, U_{XX},....)$ both the amplitudes and velocities of the ion acoustic waves and debris waves vary slowly with time. But the variation of velocity of both the wave and the debris field remain identical. Hence the ion acoustic solitary wave and the debris wave propagate simultaneously leading to ``pinned'' solitary waves as discussed in \cite{Sen}. We will consider various spatially localized functional forms of $f_1$ and determine the corresponding amplitude of the ion acoustic wave that varies slowly with time.

For the choice of  $f_1 = U$ or $U^2$,  we would finally get unforced KdV equation having standard line solitons with constant amplitudes and velocities. Thus, for that kind of choices of $f_1,$ the exact solvability of the equation will be preserved which is discussed in \cite{Sen}.  For  other  choices of $f_1$, the exact solvability of (\ref{sfKdV}) may get lost  which is discussed  in the following subsections.

\subsection{\bf $ f_1 = U_X $} 
In this case, we have considered : $F = \delta f_1 = \delta U_X$. Substituting this functional form of $F(X,T)$ and that of $U(X,T)$ from Eq. (\ref{psol}) in Eq. (\ref{condition}), we get  
\begin{equation}
\frac{dA(T)}{dT}= - (\frac{8}{15}\delta) A(T)^2, \label{dAdT1}
\end{equation}
after the integration.
In deriving this, we have used the following results of definite integration :
\begin{equation}
\int_{-\infty}^{\infty} sech^4xdx=\frac{4}{3};\,\,\int_{-\infty}^{\infty} sech^6xdx=\frac{16}{15}. \label{definite}
\end{equation} 
Thus we can see from (\ref{dAdT1}) that the amplitude of the solitary wave decays with time.
Integrating  (\ref{dAdT1})
after separation of variables, between limits $0$ and $T$, we get
\begin{equation}
\frac{A(T)}{A(0)}={(1+\frac{T}{T_0})}^{-1}; \  T_0=\frac{15}{8\delta A(0)}, \label{Decay1}
\end{equation}
where $A(0)$ is the value of $A(T)$ at $T=0$. Similar decay of amplitude was shown in \cite{OttSudan} for the case of magnetosonic waves damped by electron - ion collisions. Therefore, we would get time decaying solution for the amplitude $A(T)$, which is as shown in FIG. - 4.

\begin{figure}[hbt!]
\centering
\includegraphics[width=8cm]{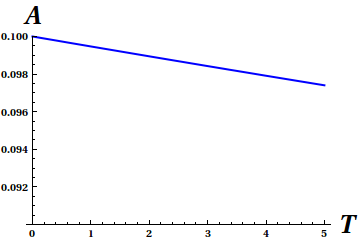}

\vspace{0.5cm}

\noindent FIG.4 :{ Decay of the solitary wave amplitude A(T) with time $T$ for Eq. (\ref{Decay1}) with $\delta = 0.1, A(0) = 0.1$.}
\label{A1}
\end{figure}

 Now, putting the expression of $A(T)$ from the above equation (\ref{Decay1}) in equation (\ref{psol}), the solitary wave solution  becomes 
\begin{equation}
 U(X, T) = A(0){(1+\frac{T}{T_0})}^{-1} \ sech^2 \ [\sqrt{\frac{A(0)}{2}}{(1+\frac{T}{T_0})}^{-1/2}\{ X - 2 \ A(0){(1+\frac{T}{T_0})}^{-1}  T \} ].\label{Decaysol1}
\end{equation}
So the amplitude, phase and velocity of the solitary wave solution
 vary with time  as seen in Eq. (\ref{Decaysol1}), showing decelerating features. In this case, the amplitude of the solitary wave decreases with time; hence the wave damps during propagation. On the other hand, for the choice : $f_1 = - U_X$, the amplitude will increase with time.

\subsection{\bf $ f_1(X, T) =   U_X^2 $} 
In this case, we have considered $F(X,T) = \delta f_1 =  \delta U_X^2$. Following the same procedure we would get,  
\begin{equation}
\frac{dA(T)}{dT}= 0. \label{dAdT2}
\end{equation}
Hence, the amplitude does not vary with time leading to an exact solitary wave solution. Similar case arises if we consider $f_1 = U^3.$ It can be shown that inclusion of these terms will transform Eq. (\ref{sfKdV}) into an extended KdV equation having exact solution \cite{ExtKdV}. Thus the amplitude and velocity of the ion acoustic solitary wave remains constant.

\subsection{ $ f_1(X, T) =   U_X^3 $} 
In this case, we have considered the cubic function, i.e $F(X,T) = \delta f_1 =  \delta U_X^3$. Following the same procedure we  get,  
\begin{equation}
\frac{dA(T)}{dT}= - (0.11 \  \delta ) \  A(T)^5. \label{dAdT3}
\end{equation}
Finally  we would get after integration :
\begin{equation}
\frac{A(T)}{A(0)}={(1+\frac{T}{T_0})}^{-1/4};  T_0=\frac{1}{0.44 \ \delta A(0)^4}, \label{Decay2}
\end{equation}
where $A(0)$ is the value of $A(T)$ at $T=0$ as in the previous case. Therefore, we get time decaying solution for the amplitude $A(T)$, which is as shown in FIG. 5.
\begin{figure}[hbt!]
\centering
\includegraphics[width=7cm]{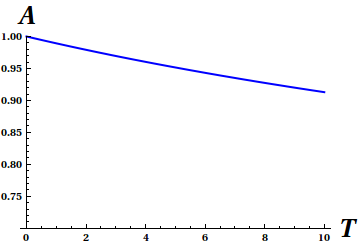}

\vspace{0.5cm}

\noindent FIG.5 :{ Decay of the solitary wave amplitude A(T) with time $T$ for Eq. (\ref{Decay2}) with $\delta = 0.1, A(0) = 1$.}
\label{A3}
\end{figure}

In the same way we can find the time variation of the amplitude of the ion acoustic solitary waves for different wave dependent localized charged debris functions. For mathematical simplicity, we have assumed polynomial dependence and found analytical expression of amplitude. For more complicated dependence of the charged debris function on the ion acoustic wave $U$, numerical calculation should be implemented.

 \section{Discussions : possible applications}

 We can sum-up our whole analytical treatment and point out few possible applications of our approach in the theoretical as well as experimental research in this field. The discussions are as follows :
    
\begin{enumerate}
\item
In this work, we have tried to analyse different exact as well as approximate ion acoustic solitary wave solutions  in presence of wave dependent localized debris functions. If any localized debris field $F$ is expressed in terms of the acoustic wave $U$, then it's analytical form may be evaluated using these discussed results.

 \item
  According to the 2015 - NASA Technology roadmap \cite{Nasa1},  ``debris
as small as 0.2 millimeter poses a realistic threat to
human spaceflight and robotic missions in the near-Earth
environment''. Hence it is necessary, to build a convincing mathematical model on the  space debris dynamics and it's detection based on more realistic nonlinear theory. A pathbreaking work on the subject of charged space debris detection in the field of nonlinear waves
has been done by
Sen et al \cite{Sen}. They  used the idea of precursor solitons for the detection of debris waves. We know that solitons are special solitary waves obeying elastic scattering property and IST solvability, where the nonlinear and
dispersive effects are balanced. Hence, the
wave propagates without dissipation preserving it's waveform. Preliminary calculations show that small orbital charged
debris ( $\sim $ cm-scale) may produce solitons \cite{Sen}
which may be detected using a rather simple
instrumentation i.e. Langmuir probe (DICE cubesat) \cite{Nasa1}. These
cubesats are used to  detect
plasma solitons that could  be used to identify meteoroids
near other planetary bodies.

\item
 Solutions to the fKdV equation take a variety of forms depending on the form  of the external disturbance. It is  observed that precursor solitons are produced when the velocity of the
source is 1-1.5 times the ion acoustic velocity \citep{Sen1, Sen2, Sen3}. For  certain forcing terms, analytical solutions (pinned solitons) can
be described. These solitons travel at the same velocity as the source.
 In realistic situations in LEO plasma region, the debris number is very large (increasing day by day) with their sizes ranging from microns to centimeters. They are also subject to many external conditions and also natural phenomena occurring in space. So there are chances that the ion acoustic solitary waves suffer frequent changes in velocities, amplitudes and phases, as we approximated here.
  Also, the change in  amplitude, velocity, phase of the solitary wave may be proportional to that in the debris function.

 \item
 
The most important and novel findings of this work are the exact accelerated solitons of the ion acoustic waves and the charged space debris functions. As far as our knowledge goes, the concept of 'exact accelerated solitons' has not been discussed in the literature of plasma physics. When the forcing function $F$ present in the forced KdV equation (\ref{ffKdV}) satisfies the constraint (\ref{NHC}), the system becomes completely integrable possessing exact accelerated solitons. The velocity of the ion acoustic soliton varies with time whereas it's amplitude remains constant. On the other way both the amplitude and velocity of the debris soliton changes causing shape changing effects. Thus this change in the amplitudes of the debris solitons may be detected and may pave a new way of debris detection. One point should be noted that both of these exact accelerated ion acoustic solitons and the debris solitons are ``pinned solitons'' that move simultaneously.  

Generally any perturbation
to an integrable system spoils it's integrability and exact solvability. But in case of integrable nonholonomic deformation theory which is applied here, the forced system retains it's integrability. Then the forcing function gets related to the ion acoustic wave in a self consistent manner that the main equation (\ref{ffKdV}) becomes integrable obeying elastic collision property and the debris function follows the nonholonomic constraint (\ref{NHC}). Physically it means that both the functions $U$ and $F$ are related to each other in such a way that the system retains it's integrable structures. On the other hand, the beauty of the exact accelerated solitons are their arbitrary acceleration. This arbitrary function $a(T)$ can be chosen accordingly to model the particular real physical condition. Generally in actual space plasma environment, the density localization may not be always of the form of line solitons. It may bend, twist, turn during it's propagation. Hence to model such real physical phenomena, soliton solutions with arbitrary free functions may be useful. In these regards, our exact accelerated solitons may be appropriate for such mathematical modelling.

\item
In an important work \cite{Kukilov}, a technique of debris detection has been discussed when the forcing functions are proportional to the nonlinear term or dispersive term of the forced KdV equation. Then those forcing parameters would be reflected in the solitary wave solutions of the system.
In our work we have used the equation :
\begin{equation}
\frac{\partial}{\partial T} \int_{-\infty}^{\infty}  U^2 dX
= \int_{-\infty}^{\infty} 2 U F_X  dX, \label{condition2}
\end{equation}
to find the amplitude variation of the ion acoustic solitary waves. In real physical situation, there may be a number of debris waves like $F_1, F_2, ....., F_N$. We can carry out similar procedure, by incorporating these weak debris functions in (\ref{condition2}) as
\begin{equation}
\frac{\partial}{\partial T} \int_{-\infty}^{\infty}  U^2 dX
= \int_{-\infty}^{\infty} 2 U \delta [ f_1 + f_2 + .... + f_N ]_X  dX, \label{condition2}
\end{equation}
where $F_i = \delta f_i$. One point should be noted that, the amplitude can be found using (\ref{condition2}) if $f_i$ is a localized function of $U$ and it's derivatives. If that functional dependence is polynomial nature, it can be solved easily as discussed before. If the dependence is more complicated, numerical procedure should be implemented. In this way we can find out the nature of solitary wave solution in presence of $N$ number of debris functions.

\item

For various localized debris functions, we can obtain both decaying as well as growing solitary wave solutions. There can also be many localized debris functions for which no time dependence of the amplitudes are resulted. So there can be a saturation in the ion acoustic wave amplitude for a large number of debris functions with different polarities and expressions.

\item
In this work, we have considered the simple physical set-up by considering propagation of ion acoustic wave in unmagnetized plasma. More interesting and complicated situation will arise if we take in-homogeneous, magnetized plasma with the inclusion of transverse effect. Obviously, then the evolution equation would be different with more complicated solutions. It would be interesting to see the effect of debris function on such complexity, that we would plan to explore in future.

\end{enumerate}

\section{Conclusive remarks}
We can conclude as, we have obtained the evolution equation for the nonlinear ion acoustic wave field induced by the orbiting charged space debris in the plasma environment generated at Low Earth Orbital (LEO)
region. The generated nonlinear ion acoustic wave is shown to be governed by the forced Korteweg-de Vries (fKdV) equation with the forcing function dependent on the charged space debris. 
In earlier scientific literatures in this field,  either numerical or perturbative solitary wave solutions have been discussed. Exact solutions are discussed in \cite{Sen} which are line solitons with constant amplitude and velocity.
 But in real physical conditions where every physical parameter changes, the localized ion density pattern also changes. To model this physical system, it is inaccurate to use line
solitons having constant amplitude and velocity. This is because the localized density structures may bend, twist or turn in $X-T$ plane depending on
physical circumstances. Hence a solitonic form with variable parameters is necessary to model such system.
For a specific self consistent condition between the ion acoustic wave $U$ and charged debris function $F$, the evolution equation turns into an integrable system where the function $F$ follows a nonholonomic constraint. 
As is well known that any forcing function usually spoils the integrability forbidding general analytic solution. However, our equation (\ref{ffKdV}) retains its integrability
in spite of the presence forcing function. This is because of the fact that
  a self consistent solitonic forcing compensates the usual loss of energy inevitable in such system. The evolution equation (\ref{ffKdV}) is solved to yield special exact accelerated soliton.
  The velocity of the ion acoustic soliton
changes over time showing accelerating/decelerating features whereas its amplitude remains constant. Whereas both the amplitude and velocity of the solitonic debris term
changes over time showing shape changing effects which may pave a new way of detection of charged debris.
For other spatially localized forms of wave dependent debris functions, the approximate solitary wave solutions are found where the amplitude varies slowly with time. Applicability of our analytical approach in various physical circumstances is also discussed.
Hence, our exact as well as approximate results may be useful for building more accurate mathematical model in this subject.

\section {ACKNOWLEDGMENTS}
The work has been carried out with partial financial support from the
Ministry of Science and Higher Education of the Russian Federation in the framework of Increase Competitiveness 
Program of NUST MISiS $( K4-2018-061),$ implemented by a governmental decree dated 16 th of March 2013, N 211. Abhik Mukherjee gratefully acknowledges late Prof. Anjan Kundu for the scintillating discussions in this field. Abhik Mukherjee would also like to thank Ms. Aurna Kar for checking the grammar and literary aspects of the manuscript.


\end{document}